\documentclass[twocolumn,showpacs,preprintnumbers]{revtex4}
\usepackage{amssymb}
\usepackage{amsfonts}
\usepackage{amsmath}
\usepackage{graphicx}
\usepackage{dcolumn}
\usepackage{bm}

\input{tcilatex}

\begin{document}

\title{Quasithermodynamic Representation of the quantum master equations:
its existence , advantages and applications}
\author{E. D. Vol}
\email{vol@ilt.kharkov.ua}
\affiliation{B. Verkin Institute for Low Temperature Physics and Engineering of the
National Academy of Sciences of Ukraine 47, Lenin Ave., Kharkov 61103,
Ukraine.}
\date{\today }

\begin{abstract}
We propose a new representation for several quantum master equations in
so-called quasithermodynamic form. This representation (when it exists) let
one to write down dynamical equations both for diagonal and nondiagonal
elements of density matrix of the quantum system of interest in unified form
by means of nonequilibrium potential ("entropy") that is a certain quadratic
function depending on all variables describing the state. We prove that
above representation exists for the general Pauli master equation and for
the Lindblad master equation ( at least in simple cases ) as well. We
discuss also advantages of the representation proposed in the study of
kinetic properties of open quantum systems particularly of its relaxation to
the stationary state.
\end{abstract}

\maketitle

\section{Introduction}

The dynamic equations method is the powerful mathematical tool in the study
of a behavior of various complex systems and therefore widely used in
physics, chemistry, population biology, and other sciences. This method may
be successfully applied both to deterministic systems (as in the case of
Newtonian mechanics where it initially arose ) and for probabilistic
description of physical and nonphysical systems both of classical and
quantum nature. For example in quantum theory of open systems (QTOS) and in
quantum optics extensively used the method of quantum master equations that
describes the evolution through time of density matrix of a system we are
interested in.It should be noted that as a rule two large classes of
dynamical systems are considered more detail in literature:1) conservative
systems and 2) dissipative ones. The first class is described by the
equations of the Hamiltonian or Lagrangian type and the second one by
equations of gradient type. In both cases there is a single function of a
system state (the Hamiltonian or Lagrangian function in the first case and
the Lyapunov(dissipative) function in the second case that completely
determines its futher evolution. In the present paper we should like to draw
attention to another important class of dynamical systems the evolution of
which is determined by two independent functions of its state. It should be
emphazised here that yet in 1865 R. Clausius one of the fathers of classical
thermodynamics had formulated \cite{1s} its two basic laws in the following
lapidar form:

\begin{equation}
\QATOP{\text{I.Die Energie der Welt ist constant}}{\text{II.Die Entropie der
Welt strebt einem Maximum zu}}  \label{c1}
\end{equation}%
It turns out that classical thermodynamics is although very important but
not the only example of similar systems.

In the paper \cite{2s}\ it was proposed to call the systems that satisfy the
two conditions Eq. (\ref{c1}) as quasithermodynamic (QT) systems .Note that
for various dynamical systems including numerous systems of nonphysical
nature the terms "energy" and "entropy" \ might be \ understood in fact only
in Pickwick sense as certain labels for two functions \ \ that satisfy to
the Clausius conditions Eq. (\ref{c1}). The main goal of present paper is to
demonstrate that certain classes of well- known in physics Markov master
equations describe substantially \ the systems of QT nature in which
diagonal $\rho _{ii\text{ }}$and nondiagonal \ $\rho _{ik}$ elements of
density matrix represent the set of dynamical variables. The pecularity of
these systems mainly consists in the fact that \ the "energy" conservation
in this case reduces merely to the standard normalization condition:\ $%
\sum\limits_{i}$\ $\rho _{ii}=1$. In such situation only one non-trivial
task remains open namely to find the explicit form of entropy function that
generates the required master equation. Of course the natural question
arises here : what advantages such QT representation \ provides one compared
with ordinary dynamical approach and surely we will\ discuss this question
in this paper.

The paper organized as follows. In Section 1 we outline those facts and \
information about QT\ systems that are necessary for the understanding of
the remainder part of the paper. In Section 2 (which is the central part of
the paper) we consider the Pauli master equation and show initially at the
simple examples and after that in general case this equation admits the
requied QT representation. We discuss also the advantages of QT
representation in the study of kinetic properties of quantum open systems
particularly in the study of the character of their relaxation to stationary
state.Also in this section we consider the case of the Lindblad master
equation that widely applyed for the description of Markov quantum systems
and show that in simple cases it admits QT representation as well.In the
Section 3 we consider more special but curious issue closely connected with
QT reperesentation namely : how under such representation "entropy" of
composite Markov system can be expressed by means of entropies of its
subsystems. We demonstrate that even in the case of noninterecting
subsystems this link turns out to be nonadditive. It should be noted however
that important issue affected in this part are needed more detail
elaboration. Now let us pass to the presentation of the results .

\section{Preliminary information}

Let us start our brief account of the theory of QT systems with the simplest
example of the dynamical system that is described by two variables $%
x_{1},x_{2\text{ \ \ \ }}$and assume it obeys the follows system of
equations :%
\begin{equation}
\frac{dx_{i}}{dt}=\varepsilon _{ik}\frac{\partial H}{\partial x_{k}}\left\{
S,H\right\} .  \label{c3am}
\end{equation}%
In Eq. (\ref{c3am})$\ H\left( x_{1},x_{2}\right) $ and $S\left(
x_{1},x_{2}\right) $ are two fixed functions of a state $\left(
x_{1},x_{2}\right) $ of the system , $\varepsilon _{ik}$ is standard
asymmetric tensor of the second rank and $\left\{ f,g\right\} $ is the
Poisson bracket for two functions $f\left( x_{1},x_{2}\right) $ and $g\left(
x_{1},x_{2}\right) $ that is $\left\{ f,g\right\} =\frac{\partial f}{%
\partial x_{1}}\frac{\partial g}{\partial x_{2}}-\frac{\partial f}{\partial
x_{2}}\frac{\partial g}{\partial x_{1}}$. It is easy to verify that
equations of motion Eq. (\ref{c3am}) imply the required two relations: 1) $%
\frac{dH}{dt}=0$ and 2)$\frac{dS}{dt}=\left\{ S,H\right\} ^{2}\geqslant 0$.
Thus the functions $H$ and $S$ satisfy to the Clausius conditions I) and II)
and hence can be considered as "energy" and "entropy" of corresponding QT.
Similarily one can consider the QT system with three variables $%
x_{1},x_{2},x_{3}$ the equations of motion of which have the form:%
\begin{equation}
\frac{dx_{i}}{dt}=\varepsilon _{ikl}\frac{\partial H}{\partial x_{k}}A_{l},
\label{c3b}
\end{equation}%
where the vector $A_{l}\equiv \varepsilon _{lmn}\frac{\partial S}{\partial
x_{m}}\frac{\partial H}{\partial x_{n}}$ and $\varepsilon _{ikl}$ is
completely antisymmetric tensor of the third rank. The equation Eq. (\ref%
{c3b}) can be written down also in the equivalent form:%
\begin{equation}
\frac{dx_{i}}{dt}=\frac{\partial S}{\partial x_{i}}\sum\limits_{k}\left( 
\frac{\partial H}{\partial x_{k}}\right) ^{2}-\frac{\partial H}{\partial
x_{i}}\sum\limits_{k}\left( \frac{\partial H}{\partial x_{k}}\frac{\partial S%
}{\partial x_{k}}\right) .  \label{c4}
\end{equation}%
It should be noted however that Eq. (\ref{c3b}) and Eq. (\ref{c4}) are not
the most general form of equations for QT systems with three variables. In
fact we may add in r.h.s. of the equation Eq. (\ref{c3b}) the hamilton- like
term- $r\varepsilon _{ikl}\frac{\partial S}{\partial x_{k}}\frac{\partial H}{%
\partial x_{l}}$ (where r is an arbitrary multiplier) without changing its
QT nature. So the general form of QS with three variables may be written as:%
\begin{equation}
\frac{dx_{i}}{dt}=\epsilon _{ikl}\frac{\partial H}{\partial x_{k}}\left(
A_{l}-r\frac{\partial S}{\partial x_{l}}\right) .  \label{c5}
\end{equation}%
The construction of the explicit form of equations of motion for QT systems
with more than three variables can be solved in principle by the similar way
and we will return to it in the next part where the general Paulu master
equation would be considered as the example of QT system. Now we briefly
explain: why for probabilistic systems the equations of motion in many cases
admit QT representation. Indeed let us assume that probabilistic system of
interest is described by dynamical equations that have the following
schematic form:%
\begin{equation}
\frac{dp_{i}}{dt}=F_{i}\left\{ p_{\alpha }\right\} .  \label{c6m}
\end{equation}%
In Eq. (\ref{c6m}) $p_{i}$ is a probability to detect the system in the
state $i$ ($i=1,2...N)$. The above formulation of the problem assumes that
normalization condition $\sum\limits_{i=1}^{N}p_{i}=1$ and hence the
restriction $\sum\limits_{i=1}^{N}F_{i}=0$ is satisfied. Thus we conclude
that for probabilistic QT systems one of the two functions that determine
such systems namely energy must be defined as $H=\sum\limits_{i=1}^{N}p_{i}$%
. In respect of the entropy function its existence and concrete form should
be established in every individual case separately.

\section{The Pauli master equation and its QT representation}

Now we consider one extensive class of probabilistic systems namely those
which behavior can be described by the Pauli-master equation (PME) $\left[ 3%
\right] $.It is well known the PME describes the evolution through time only
diagonal elements of density matrix of N-state open quantum system of
interest and may be written down in the next standard form:

\begin{equation}
\frac{dP_{i}}{dt}=\sum\limits_{k=1}^{N}\left( W_{ik}P_{k}-P_{i}W_{ki}\right)
.  \label{c7m}
\end{equation}%
In Eq. (\ref{c7m}) $P_{i}\left( t\right) $ is a probability to detect the
system of interest in quantum state $\left\vert i\right\rangle $ , $W_{ik}$
is a probability (per unit time) of transition from state $\left\vert
k\right\rangle $ to state $\left\vert i\right\rangle $. It is assumed the
set of states $\left\{ \left\vert i\right\rangle \right\} $ $\left(
i=1...N\right) $ forms some complete basis in N state vector space.
Comparing Eq. (\ref{c6m}) and Eq. (\ref{c7m}) we conclude that in the case
of the PME $F_{i}\left\{ P_{\alpha }\right\} \equiv
\sum\limits_{k=1}^{N}\left( W_{ik}P_{k}-P_{i}W_{ki}\right) $ and the
necessary restriction $\sum F_{i}=0$ is satisfied automatically. Note that
kinetic properties of the system described by the Eq.(7) to a large extent
depend on the restrictions imposed on the coefficients $\ W_{ik}$. In his
prominent paper \cite{4s} J.S. Tomsen obtained some important relations
connecting symmetry properties of coefficients $W_{ik}$ with the character
of relaxation process in corresponding quantum system obeying the PME.\ For
example if these coefficients are symmetric that is $W_{ik}=W_{ki\text{ }}$
then final probabilities $\left\{ P_{i}^{0}\right\} $ to find the system in
its stationary state are identical i.e. the ergodic hypothesis in this case
is true. On the other hand the more weak property of matrix $W_{ik}$ namely
its double stochasticity : $\sum\limits_{k}W_{ik}=\sum\limits_{k}W_{ki}$ for
all indexes $i$ implies the Boltzmann-Shannon entropy function $%
S_{BS}=-\sum\limits_{i}P_{i}\ln P_{i}$ increases in time($\frac{dS}{dt}%
\geqslant 0)$.Therefore we believe that in symmetric case the PME in fact
describes the evolution of the quantum system of interest to its equilibrium
state.Note that in present paper we do not assume (until contrary is not
approved ) any special symmetry properties of coefficients $W_{ik.}$

Let us begin our study with the simplest case of two state quantum system
described by the PME.We write down the PME for the diagonal matrix elements
of its density matrix $\widehat{\rho }$, namely $p_{1}\equiv \rho _{11\text{ 
}}$and $p_{2}\equiv \rho _{22}$ as:%
\begin{equation}
\QATOP{\frac{dp_{1}}{dt}=W_{12}p_{2}-p_{1}W_{21},}{\frac{dp_{2}}{dt}%
=W_{21}p_{1}-p_{2}W_{12}.}  \label{c8b}
\end{equation}%
One can directly verify that the system Eq. (\ref{c8b}) may be represented
in required QT form : $\frac{dp_{i}}{dt}=\varepsilon _{ik}\frac{\partial H}{%
\partial p_{k}}\left\{ S,H\right\} $ if one takes the "energy" function as $%
H=p_{1}+p_{2}$ and "entropy" function as $S=-\frac{W_{12}p_{1}^{2}}{2}-\frac{%
W_{21}p_{2}^{2}}{2}$. It should be noted if the symmetry condition $%
W_{12}=W_{21}$ is realized the above entropy function in fact coincides with
linear Boltzmann-Shannon entropy that provides the relaxation of the system
to its equilibrium state with $p_{1}^{0}=p_{2}^{0}=\frac{1}{2}$ .But for
general two state Markov system we obtain the stationary probabilities as : $%
p_{1}^{0}=\frac{W_{12}}{W_{12}+W_{21}}$ and $p_{2}^{0}=\frac{W_{21}}{%
W_{12}+W_{21}}$ and ergodic hypothesis does not hold.It is clear that two
state case is too simple to shed light on behavior of general $N$ state
Markov system but in the next in complexity three-state case which admits
the complete inquiry as well all key elements of required general
construction can be recognized. So let us consider the case of three-state
Markov system more detail. The dynamical equations for such system can be
written in the following form%
\begin{equation}
\begin{array}{c}
\frac{dp_{1}}{dt}=-\left( a+b\right) p_{1}+cp_{2}+ep_{3}, \\ 
\frac{dp_{2}}{dt}=ap_{1}-\left( c+d\right) p_{2}+fp_{3}, \\ 
\frac{dp_{3}}{dt}=bp_{1}+dp_{2}-\left( e+f\right) p_{3}.%
\end{array}
\label{c9c}
\end{equation}%
It is clear that there is complete coincidance between the PME Eq. (\ref{c7m}%
) in the case when $N=3$ and the system Eq. (\ref{c9c}). To this end enough
to identify $a$ with $W_{21\text{, }}b$ with $W_{31},c$ with $W_{12}$, $d$
with $W_{32}$, $e$ with $W_{13\text{ }}$and $f$ with $W_{23}$. Note that in
the situation of general $N$-state PME one has obviously $N\left( N-1\right) 
$ independent coefficients in it and hence in the three state case there are
precisely six free parameters. Now let us seek the desired representation of
the system Eq. (\ref{c9c}) in the required QT form as%
\begin{equation}
\frac{dp_{i}}{dt}=\varepsilon _{ikl}\frac{\partial H}{\partial p_{k}}\left(
A_{l}-r\frac{\partial S}{\partial p_{l}}\right) ,  \label{c10}
\end{equation}%
where all indices take values 1, 2, 3, and according to definition the
vector $A_{l}=\varepsilon _{lmn}\frac{\partial S}{\partial p_{m}}\frac{%
\partial H}{\partial p_{n}}$, $H=p_{1}+p_{2}+p_{3}$, and $r$ is scalar
factor. Note that in fact Eq. (\ref{c10})\ coincides with Eq. (\ref{c5}) but
with concrete energy function.

As regards to the "entropy" function $\ S$ we will seek it in the form of
arbitrary quadratic function of basic variables $p_{i}$%
\begin{equation}
S=\frac{Ap_{1}^{2}}{2}+\frac{Bp_{2}^{2}}{2}+\frac{Cp_{3}^{2}}{2}+\alpha
p_{1}p_{2}+\beta p_{1}p_{3}+\gamma p_{2}p_{3}.  \label{c11}
\end{equation}%
It is easy to see that the transformation $S\rightarrow S+k\left(
p_{1}+p_{2}+p_{3}\right) ^{2}$ does not change equations of motion Eq. (\ref%
{c10}) and therefore without loss of generality one can put the value of $%
\gamma $ to be equal zero. Substituting the expression Eq. (\ref{c11}) into
Eq. (\ref{c10}) and comparing the coefficients of identical powers of $p_{1%
\text{ ,\ \ \ }}p_{2\text{ \ , }}p_{3\text{ }}$ with Eq. (\ref{c9c}) one can
find all unknown parameters $A$, $B$, $C$, $\alpha $, $\beta $\ and
reconstruct QT representation of the PME Eq. (\ref{c9c}) in explicit form.
We adduce here only the expression of the parameter $r$ that does not depend
on the concrete choice of the entropy function: $r\equiv \frac{1-\varkappa }{%
1+\varkappa }$, where $\varkappa =\frac{b+c+f}{a+d+e}$. It turns out and
this fact is the instructive argument in behalf of QT representation that
the condition $r=0$ results in to monotonic relaxation of the system of
interest to its stationary state. Let us prove this statement now. Indeed we
can seek the solutions of linear PME Eq. (\ref{c9c}) in standard form as $%
p_{i}\left( t\right) =C_{i}e^{\lambda t}$ and after a simple algebra we
obtain the qubic secular equation for its three roots. One root is precisely
equal to zero since the sum $\sum\limits_{i=1}^{i=3}p_{i}$ is conserved. The
other two roots can be obtained from the following quadratic equation:%
\begin{equation}
\lambda ^{2}+\xi \lambda +\eta \left( a+b+e\right) -\left( e-c\right) \left(
f-a\right) =0,  \label{c12}
\end{equation}%
where, $\xi =a+b+c+d+e+f$, $\eta =c+d+f$. The condition that the determinant
of this equation less than zero implies two roots of Eq. (\ref{c12}) will be
real and negative. Thus the necessary and sufficient condition of monotonic
relaxation of open Markov system Eq. $\left( 9\right) $ to its stationary
state may be written as:%
\begin{equation}
\xi ^{2}+4\left( e-c\right) \left( f-a\right) -4\eta \left( a+b=e\right)
\leqslant 0.  \label{c13}
\end{equation}%
Let us use the notation: $k=e-c,l=f-a,m=b-d$ and $\omega =\left(
a+d+e\right) -\left( b+c+f\right) .$ In this notation the condition Eq. (\ref%
{c13}) looks as $\ \omega ^{2}+4\omega \left( l+m\right) +4\left(
l^{2}+m^{2}+lm\right) \leqslant 0$ $\ $or in more convinient form as $\left( 
\sqrt{3}u+\frac{2}{\sqrt{3}}\omega \right) ^{2}+v^{2}-\frac{\omega ^{2}}{3}%
\leqslant 0$ \ where $u\equiv l+m$ and $v\equiv l-m$. We see that the
boundary of the region in parameter space of the PME Eq. (\ref{c9c}), where
the nonmonotonic relaxation of its solutions is possible may be represented
by the ellipse: $\left( \sqrt{3}u+\frac{2}{\sqrt{3}}\omega \right)
^{2}+v^{2}=\frac{\omega ^{2}}{3}$. Obviously when $\omega =0,$ i.e. the
condition $a+d+e=b+c+f$ \ or \ $r=0$ holds, the ellipse degenerates into
single point and all solutions of Eq. (\ref{c9c}) monotonically decrese in
time. On the other hand if $\omega \neq 0$ there is a finite region of
parameters (the greater the more $\omega $ is) where nonmonotonic behavior
of solutions of Eq. (\ref{c9c}) is possible. So the result stated above is
proved. Now let us prove the existence of QT representation of the PME in
general case of $N$ state open quantum system. It should be noted that the
construction of QT representation of the Eq. (\ref{c9c}) may be realized
with necessary changes in general case as well. We propose here only the
outline of a complete proof. So let us consider the $N$ state Markov system
described by corresponding PME Eq. (\ref{c7m}) with $N\left( N-1\right) $
independent coefficients.We claim that required QT representation of this
PME may be represented in the next form:%
\begin{equation}
\frac{dp_{i}}{dt}=\varepsilon _{i,i_{1}...i_{N-1}}\frac{\partial H}{\partial
p_{i_{1}}}A_{i_{2}...i_{N-1}}+\sum\limits_{\alpha =1}^{M}r_{\alpha
}H_{i}^{\left( \alpha \right) },  \label{c14}
\end{equation}%
where $M=\frac{\left( N-1\right) \left( N-2\right) }{2}$, $p_{i}$ has the
same sense as in the Eq. (\ref{c6m}), $H=p_{1}+....p_{N}$ , $%
A_{i_{2}.....i_{N-1}}=\varepsilon _{i,i_{1}.....i_{N-1}}\frac{\partial S}{%
\partial p_{i}}\frac{\partial H}{\partial p_{i_{1}}}$ $\left( \varepsilon
_{i_{1}}..._{i_{N}}\text{ is completely antisymmetric tensor of }N\text{ rank%
}\right) $ and each of the $\frac{\left( N-1\right) \left( N-2\right) }{2}$
hamiltonian like terms $H_{i}^{\left( \alpha \right) }$ can be constructed
by the following procedure. Firstly let us consider in $N$ dimensional
vector space representing all states of the system the subspace (hyperplane)
consisting of all states that are orthogonal to the vector $\frac{\partial H%
}{\partial p_{i}}=\left( 1,1...1\right) .$ Obviously this hyperplane has
dimensionality $N-1$. Further we choose from the basis of this hyperplane
arbitrarily $N-3$ vectors and construct by standard way on these vectors the
antisymmetric tensor of $N-3$ rank. Each of this tensors (with accompanying
coefficient $r_{\alpha }$ ) enters in the sum in r.h.s. of Eq. (\ref{c14}).
It is clear that we obtain in this way precisely $C_{N-1}^{N-3}=C_{N-1}^{2}$
distinct hamiltonian- like terms and respectively $C_{N-1}^{2}$ free
parameters $r_{\alpha }$. Let us calculate now the total number of free
parameters being in our disposal. The entropy function as symmetric
quadratic form of N variables gives us $\left[ \frac{N\left( N+1\right) }{2}%
-1\right] $ parameters (we take here into account that $S$ is defined up to
the term $k\left( p_{1}+....p_{N}\right) ^{2}$ ). Besides due to various
choice of hamiltonian- like terms we get additional $C_{N-1}^{2}$
parameters.Thus as the final result we have $\frac{N\left( N+1\right) }{2}-1+%
\frac{\left( N-1\right) \left( N-2\right) }{2}=N\left( N-1\right) $ free
parameters that is as much as we need for the initial PME Eq. (\ref{c7m}).
This simple reasoning in our opinion proves our original statement. . Now we
demonstrate that the PME is not the only quantum master equation which
admits QT representation.In particular well-known Lindblad master equation
(LME) that describes the evolution of arbitrary quantum Markov system admits
the similar QT representation at least in the special case of two state
systems as well. So let us consider this special case more thoroughly.We
start with general Lindblad master equation which has the following form 
\cite{3s}%
\begin{equation}
\frac{d\rho }{dt}=-\frac{i}{\hbar }\left[ H,\rho \right] +\sum%
\limits_{j=1}^{N}\left[ R_{j}\rho ,R_{j}^{+}\right] +h.c,  \label{c15}
\end{equation}%
(where $H$ some hermitian operator,describing intrinsic dynamics of the open
system and operators $\left\{ R_{j}\right\} $ are the set of nonhermitian
operators that describe the interaction of the system of interest with
environment.In the case of two state \ open systems that we are only
interested in here it is convenient to use the Bloch representation for its
density matrix,namely $\rho =\frac{1+\overrightarrow{P}\text{ }%
\overrightarrow{\sigma }}{2}$, where $\overrightarrow{P}$ is polarization
vector of the state and $\overrightarrow{\sigma }=\left\{ \sigma
_{k}\right\} (k=1,2,3)$ are standard Pauli matrices. Taking into account
that any $2\times 2$ hermitian matrix can be decomposed in Pauli matrices
one can write down all operators entering in Eq. (\ref{c15}) in the
following form: $H=2\overrightarrow{h}$ $\overrightarrow{\sigma },$ and $%
R_{j}=\overrightarrow{A_{j}}$ $\overrightarrow{\sigma }+i\overrightarrow{%
B_{j}}$ $\overrightarrow{\sigma }$ where $i$ $\equiv \sqrt{-1}$.The set of
vectors $\overrightarrow{h},\overrightarrow{A_{j}},\overrightarrow{B_{j}}$
are completely characterizes the evolution of two state open system within
the Lindblad equation approach. Based on the Eq. (\ref{c15}) and using the
Bloch representation of input operators we can write down the LME in two
state case as equation for the state vector $\overrightarrow{P}$, namely:%
\begin{eqnarray}
\frac{d\overrightarrow{P}}{dt} &=&\left( \overrightarrow{h}\times 
\overrightarrow{P}\right) +\sum\limits_{j=1}^{N}2\left( \overrightarrow{A_{j}%
}\times \overrightarrow{B_{j}}\right)   \label{c16} \\
&&-\overrightarrow{A_{j}}\times \left( \overrightarrow{P}\times 
\overrightarrow{A_{j}}\right) -\overrightarrow{B_{j}}\times \left( 
\overrightarrow{P_{j}}\times \overrightarrow{B_{j}}\right) .  \notag
\end{eqnarray}%
In what follows for the simplicity we will consider the special case when $%
N=1$ that is only one operator $R$ \ in the r.h.s. of Eq. (\ref{c15}) is
nonzero. In addition we assume that there is no hamiltonian- like term $%
\left( \overrightarrow{h}\times \overrightarrow{P}\right) $ in Eq. (\ref{c16}%
). After these assumptions the simplified version of Eq. (\ref{c16}) takes
the form%
\begin{eqnarray}
\frac{d\overrightarrow{P}}{dt} &=&2\left( \overrightarrow{A}\times 
\overrightarrow{B}\right) -\overrightarrow{A}\times \left( \overrightarrow{P}%
\times \overrightarrow{A}\right)   \label{c17} \\
&&-\overrightarrow{B}\times \left( \overrightarrow{P}\times \overrightarrow{A%
}\right) .  \notag
\end{eqnarray}%
Note that Eq. (\ref{c17}) implies that the Bloch vector of the stationary
state is equal to $\overrightarrow{P}_{st}=\frac{2\left( \overrightarrow{A}%
\text{ }\times \overrightarrow{B}\text{\ }\right) }{A^{2}+B^{2}}$ and hence
if one takes the operator $\widehat{R}=\overrightarrow{A}\overrightarrow{%
\sigma }+i\overrightarrow{B}\overrightarrow{\sigma }$ \ so that $\left\vert 
\overrightarrow{A}\right\vert =\left\vert \overrightarrow{B}\right\vert $ and%
$\overrightarrow{\text{ \ }A}\cdot \overrightarrow{B}=0$ the final
stationary state would be pure one irrespective of initial state of the
system. Now it is easy to verify directly that the LME in vector form Eq. (%
\ref{c17}) can be represented also in gradient form as:

\begin{equation}
P_{i}=\frac{\partial S}{\partial P_{i}}.  \label{c18mm}
\end{equation}%
\ To this end one need to choose the entropy function as%
\begin{eqnarray}
S\left( \overrightarrow{P}\right) &=&2\left( \overrightarrow{A}\times 
\overrightarrow{B}\right) \cdot \overrightarrow{P}-\frac{P^{2}}{2}\left(
A^{2}+B^{2}\right)  \label{c19} \\
&&+\frac{\left( \overrightarrow{A}\overrightarrow{P}\right) ^{2}}{2}+\frac{%
\left( \overrightarrow{B}\overrightarrow{P}\right) ^{2}}{2}.  \notag
\end{eqnarray}%
Thus we have seen that the LME Eq. (\ref{c17}) admits the representation in
simple gradient form. Now we show that specified gradient system with three
variables $P_{i}$ Eq. (\ref{c18mm}) can be in natural way represented as QT
system with 6 variables. So let $P_{x},P_{y},P_{z}$ are three components of
the Bloch vector satisfying to Eq.$\left( 18\right) .$ Then by means of this
components we introduce six new variables $\left\{ p_{i}\right\} $ $\left(
i=1...6\right) $ according to the rule: $p_{1}=\frac{1+P_{x}}{2}$, $p_{2}=%
\frac{1-P_{x}}{2}$, $p_{3}=\frac{1+P_{y}}{2}$, $p_{4}=\frac{1-P_{y}}{2}$, $%
p_{5}=\frac{1+P_{z}}{2}$, $p_{6}=\frac{1-P_{z}}{2}$. Now let us write the
following QT system of equations for variables $\ p_{i}$ :%
\begin{equation}
\frac{dp_{i}}{dt}=N\varepsilon _{iklmnp}\frac{\partial H_{1}}{\partial p_{k}}%
\frac{\partial H_{2}}{\partial p_{l}}\frac{\partial H_{3}}{\partial p_{m}}%
A_{np},  \label{c20}
\end{equation}%
where $H_{1}=p_{1}+p_{2},H_{2}=p_{3}+p_{4}$, $H_{3}=p_{5}+p_{6}$ are three
integrals of motion for the Eq. (\ref{c20}), $\varepsilon _{iklmnp\text{ }}$%
is the antisymmetric tensor of the 6 rank, and tensor $A_{np}$ according to
definition is :$A_{np}=N\varepsilon _{nprstu}\frac{\partial S}{\partial p_{r}%
}\frac{\partial H_{1}}{\partial p_{s}}\frac{\partial H_{2}}{\partial p_{t}}%
\frac{\partial H_{3}}{\partial p_{u}}$ where entropy function of three
variables $S=S\left( p_{1}-p_{2},p_{3}-p_{4},p_{5}-p_{6}\right) $ up to
notation should be coincide with entropy function Eq. (\ref{c19}), $N$ is
normalizing factor. Let us prove now that under appropriate choice of
coefficient $N$ the equations Eq. (\ref{c20}) are in fact entirely
equivalent to equations Eq. (\ref{c18mm}). We test this statement only for
the first pair of variables of Eq. (\ref{c20}) namely $p_{1\text{ \ }}{}_{%
\text{\ }}$and $p_{2}$. All other equations may be obtain in a similar way
as well. Indeed the first equation Eq. (\ref{c20}) in expended form looks as
follows:%
\begin{equation}
\frac{dp_{1}}{dt}=2N\left( A_{64}+A_{45}+A_{36}+A_{53}\right) .  \label{c21}
\end{equation}%
Calculating coefficients $A_{np}$ entering in Eq. (\ref{c21}) in explicit
form (using the above expressions for them) we obtain : $%
A_{64}=A_{45}=A_{36}=A_{53}=\frac{\partial S}{\partial p_{1}}-\frac{\partial
S}{\partial p_{2}}$ and hence $\frac{dp_{1}}{dt}=8N^{2}\left( \frac{\partial
S}{\partial p_{1}}-\frac{\partial S}{\partial p_{2}}\right) $. By selfsame
way we obtain that $\frac{dp_{2}}{dt}=8N^{2}\left( \frac{\partial S}{%
\partial p_{2}}-\frac{\partial S}{\partial p_{1}}\right) .$ Hence as a
result for component $P_{x}=p_{1}-p_{2}$ \ we obtain the equation $\frac{%
dP_{x}}{dt}=16N^{2}\left( \frac{\partial S}{\partial p_{1}}-\frac{\partial S%
}{\partial p_{2}}\right) =64N^{2}\frac{\partial S}{\partial P_{x}}$. Thus we
conclude that if one choose the coefficient $N$ as $\frac{1}{8}$ the
equations Eq. (\ref{c20}) and Eq. (\ref{c18mm}) would be entirely
equivalent.Thus we have proved that in two state case the LME admits the
required QT representation. Relating to the possibility to represent the LME
in more general situation note that this issue is rather delicate. We
believe that this possibility is in fact closely connected with the problem
of existence of hidden variables for system under consideration and
therefore goes far beyond the scope of this paper.

\section{The simple composite Markov system consisting of two independent
subsystems and subextensivity of its entropy function.}

In this part we consider one important issue closely connected with QT
representation of quantum master equations namely: how the "entropy"
function of composite quantum system satisfying to the PME may be expressed
by means of entropies of its subsystems.In present paper we study only the
simplest example of this problem namely we study the four state composite
Markov system described by the PME that consists of a pair of two state
independent subsystems . We show that even in this case required connection
turns out to be subextensive that is to a certain extent is the same as in
the nonextensive statistical thermodynamics $\left[ 6\right] $. We start
with two statistical independent quantum systems $A$ and $B$ both of which
may be described by the PME ,namely, for system $A$%
\begin{equation}
\QATOP{\frac{dp_{1}}{dt}=-ap_{1}+bp_{2},}{\frac{dp_{2}}{dt}=ap_{1}-bp_{2}}
\label{c22b}
\end{equation}%
and for system $B$%
\begin{equation}
\QATOP{\frac{dq_{1}}{dt}=-cq_{1}+dp_{2},}{\frac{dq_{2}}{dt}=cq_{1}+dq_{2}.}
\label{c23b}
\end{equation}%
In what follows we will consider only the situation when $a=b$ and $c=d$
because as we will see in this case QT representation of the PME for the
composite system $C$ consisting of subsystems $A$ and $B$ does not contain
hamiltonian like terms. Note that in this case the entropies of subsystems $%
A $ and $B$ can be chosen in the next form:$S_{A}=-\frac{a\left(
p_{1}-p_{2}\right) ^{2}}{4}$ and $S_{B}=-\frac{c\left( q_{1}-q_{2}\right)
^{2}}{4}.$ Let us introduce now the probabilities of populations $W_{\alpha
}\left( \alpha =1,..4\right) $ for the four states of composite system $C.$
In view of the statistical independence $A$ and $B$ one can express these
probabilities $W_{\alpha }$ by means of probabilities $\ p_{i}$ and $q_{k}$
as follows : $%
W_{1}=p_{1}q_{1},W_{2}=p_{1}q_{2},W_{3}=p_{2}q_{1},W_{4}=p_{2}q_{2}$. The
conditions: $p_{1}+p_{2}=1$ and $q_{1}+q_{2}=1$ imply that $%
\sum\limits_{\alpha =1}^{\alpha =4}W_{\alpha }=1.$ In addition dynamical
equations Eq. (\ref{c22b}) and Eq. (\ref{c23b}) immediately imply the
following equations of motion for probabilities $W_{\alpha }$%
\begin{equation}
\begin{array}{c}
\frac{dW_{1}}{dt}=-\left( a+c\right) W_{1}+cW_{2}+aW_{3}, \\ 
\frac{dW_{2}}{dt}=cW_{1}-\left( a+c\right) W_{2}+aW_{4}, \\ 
\frac{dW_{3}}{dt}=aW_{1}-\left( a+c\right) W_{3}+cW_{4}, \\ 
\frac{dW_{4}}{dt}=aW_{2}+cW_{3}-\left( a+c\right) W_{4}.%
\end{array}
\label{c24d}
\end{equation}%
It is obvious that Eq. (\ref{c24d}) is the PME for the composite system $C$
and therefore as it was proved above admit QT representation with some
entropy function $S$. Our task is to find explicit form of this function and
to state its connection with subsystems entropies $S_{A}$ and $\emph{S}_{B}$%
. Let us seek the required QT representation of Eq. (\ref{c24d}) in the form%
\begin{equation}
\frac{dW_{i}}{dt}=N\varepsilon _{iklm}\frac{\partial H}{\partial W_{k}}%
A_{lm}.  \label{c25}
\end{equation}%
In Eq. (\ref{c24d}) according to definition the antisymmetric tensor $%
A_{lm}=N\varepsilon _{lmnp}\frac{\partial S}{\partial W_{n}}\frac{\partial H%
}{\partial W_{p}}$ , $\varepsilon _{iklm}$ is completely antisymmetric
tensor of fourth rank, $H=\sum\limits_{i=1}^{i=4}W_{i}$, $N$ - some
normalizing factor and $S$ is the entropy function of the composite system $%
C $ which leads to the equations coinciding with Eq. (\ref{c24d}). Let us
take the entropy function $S$ in the following subextensive form : $%
S=S_{A}+S_{B}-\lambda S_{A}S_{B}$ where the factor $\lambda $ would be
specified. Taking into account that$\ S_{A}=$ $-\frac{a}{4}\left(
p_{1}-p_{2}\right) ^{2}$ $=$ $-\frac{a}{4}\left(
W_{1}+W_{2}-W_{3}-W_{4}\right) ^{2}$, $S_{B}=-\frac{c}{4}\left(
q_{1}-q_{2}\right) ^{2}$ $=$ $-\frac{c}{4}\left(
W_{1}+W_{3}-W_{2}-W_{4}\right) ^{2}$ and $S_{A}S_{B}$ $=$ $\frac{ac}{16}%
\left( W_{1}+W_{4}-W_{2}-W_{3}\right) ^{2}$ one can obtain the expression
for the entropy function $S$ as:%
\begin{eqnarray}
S &=&-\frac{a}{4}\left( W_{1}+W_{2}-W_{3}-W_{4}\right) ^{2}  \label{c26} \\
&&-\frac{c}{4}\left( W_{1}+W_{3}-W_{2}-W_{4}\right) ^{2}  \notag \\
&&-\lambda \frac{ac}{16}\left( W_{1}+W_{4}-W_{2}-W_{3}\right) ^{2}.  \notag
\end{eqnarray}%
We prove that by an appropriate choice of factor $\lambda $ QT
representation of composite systems that is Eq. (\ref{c25}) coincides with
the PME Eq. (\ref{c24d}). We give here the proof only for the first equation
of Eq. (\ref{c24d}), all others equations can be obtained in a similar way.
So, the first of Eq. (\ref{c24d}) in extended form reads as:%
\begin{equation}
\frac{dW_{1}}{dt}=2N\left( A_{23}+A_{34}+A_{42}\right) .  \label{c27}
\end{equation}%
According to definition all tensors $A_{lm}$ entering in Eq. (\ref{c27}) can
be easily calculated and are equal to: $A_{23}=N\left( \frac{\partial S}{%
\partial W_{1}}-\frac{\partial S}{\partial W_{4}}\right) $, $A_{34}=N\left( 
\frac{\partial S}{\partial W_{1}}-\frac{\partial S}{\partial W_{2}}\right) $
and $A_{42}=N\left( \frac{\partial S}{\partial W_{1}}-\frac{\partial S}{%
\partial W_{3}}\right) $. Substituting these expressions in Eq. (\ref{c27})
we obtain that%
\begin{equation}
\frac{dW_{1}}{dt}=2N^{2}\left( 4\frac{\partial S}{\partial W_{1}}%
-\sum\limits_{i=1}^{i=4}\frac{\partial S}{\partial W_{i}}\right) .
\label{c28}
\end{equation}%
It is easy to see that expression Eq. (\ref{c26}) implies that $%
\sum\limits_{i=1}^{i=4}\frac{\partial S}{\partial W_{i}}=0$ and if one
chooses the factor $N$ \ so that $8N^{2}=1$ the equation for probability $%
W_{1}$ takes standard dissipative form:%
\begin{equation}
\frac{dW_{1}}{dt}=\frac{\partial S}{\partial W_{1}}.  \label{c29}
\end{equation}%
Using the expression Eq. (\ref{c26}) for the entropy function of the
composite system it\ \ is not difficult to verify that if one choose the
factor $\lambda $ so that $\frac{ac\lambda }{8}=\frac{a+c}{2}$ the first
equation of Eq. (\ref{c24d})\ coincides with Eq. (\ref{c29})\ and the
required result is proved. It should be noted here that in nonextensive
thermodynamics (see i.e \cite{6s}\ ) the following general expression for
the nonextensive entropy takes place: $S_{AB}=$ $S_{A}+S_{B}+\frac{1-q}{k}%
S_{A}S_{B}$ where $k$\ is the Boltzmann constant and $q$\ \ is the factor of
nonextensivity. Note that superextensivity, extensivity and subextensivity
occurs when \ $q\leqslant 1$, $q=1$, $q\geqslant 1$ respectively.In our case 
$\left( q=1+\frac{k\left( a+c\right) }{4}\right) $\ and hence in this model
we are dealing with subextensive situation.

In conclusion let us express another general reason on behalf of possible
advantage of QT representation of quantum master equations.It is well-known
that the formulation of principles of mechanics and field theory in
Hamiltonian or Lagrangian form enables one to describe dynamics of the
complex conservative systems on the basis of knowledge their parts behavior
and the symmetry of interaction between all their parts. Similarly we
believe that the deeper understanding of QT representation of master
equations let one construct more and more complex probabilistic models based
on the collection of certain elementary constituents that admit detail
analysis.

I am very obliged to L.A. Pastur for discussions of the results of the paper
and valuable comments.

\end{document}